%% file: main.tex
\documentclass[sigconf, screen, nonacm]{acmart}
\AtBeginDocument{%
  }

\setcopyright{none}
\input{our_imports}

\begin{document}

\title{From Measurement to Mitigation: Aligning Recommendations with User Popularity Preferences}
\title{Aligning Recommendations with User Popularity Preferences}
\author{Mona Schirmer}
\authornote{Work done while interning at Amazon Music.}
\affiliation{%
  \institution{University of Amsterdam}
    \city{Amsterdam}
  \country{Netherlands}
}
\email{m.c.schirmer@uva.nl}

\author{Anton Thielmann}
\affiliation{%
  \institution{Amazon Music}
  \city{Berlin}
  \country{Germany}
}
\email{thielmaf@amazon.de}

\author{Pola Schwöbel}
\affiliation{%
  \institution{Amazon Music}
    \city{Berlin}
  \country{Germany}
}
  \email{schwobel@amazon.de}

\author{Thomas Martynec}
\affiliation{%
  \institution{Amazon Music}
    \city{Berlin}
  \country{Germany}
}
  \email{martynec@amazon.de}

\author{Giuseppe Di Benedetto}
\affiliation{%
  \institution{Amazon Music}
    \city{Berlin}
  \country{Germany}
}
  \email{bgiusep@amazon.de}

\author{Ben London}          
\affiliation{%
  \institution{Amazon Music}
    \city{Seattle}
  \country{USA}
}
  \email{blondon@amazon.com}

\author{Yannik Stein}
\affiliation{%
  \institution{Amazon Music}
    \city{Berlin}
  \country{Germany}
}
  \email{syannik@amazon.de}

\renewcommand{\shortauthors}{Schirmer et al.}


\input{text/abstract}

\ccsdesc[500]{Information systems~Recommender systems}
\ccsdesc[500]{General and reference~Metrics}

\keywords{Recommender Systems, Popularity Bias, Measurements}

\maketitle

\input{text/introduction}
\input{text/related-work}
\input{text/popularity_bias_problem}

\input{text/measuring-bias-past-work}

\input{text/measuring-bias-ours}

\input{text/mitigating-bias}

\input{text/experiments}
\input{text/conclusion}

\subsection*{Acknowledgments}
We thank Masoud Mansoury for helpful discussions on the Popsteer paper, and the anonymous reviewers for their suggestions, which helped refine this work.

\section*{Author Contribution Statement}
\noindent \textbf{Project Conception} [\textsc{Schirmer}, \textsc{Stein}, \textsc{Thielmann}, \textsc{Schwöbel}] \\
\textbf{Analysis Advisory} [\textsc{Everyone}] \\
\textbf{Measurement Theory Analysis} [\textsc{Schirmer}, \textsc{Schwöbel}] \\
\textbf{Calibration Metric} [\textsc{Schirmer}, \textsc{Martynec}] \\
\textbf{Steering Methodology} [\textsc{Schirmer}] \\
\textbf{Experiments} [\textsc{Schirmer}, \textsc{Thielmann}] \\
\textbf{Literature} [\textsc{Schirmer}, \textsc{Di Benedetto}, \textsc{Schwöbel}] \\
\textbf{Writing} [\textsc{Schirmer}, \textsc{London}, \textsc{Schwöbel}, \textsc{Di Benedetto}, \textsc{Stein}]

\subsection*{Generative AI Usage Statement}
We used ChatGPT and Claude for grammar correction, LaTeX compilation, and assistance with plotting. It was not used to generate manuscript text or scientific content from scratch. The authors retain full responsibility for the originality, accuracy, and integrity of the manuscript.


\bibliographystyle{ACM-Reference-Format}
\bibliography{bib}


\newpage
\begin{appendices}
\appendix
\crefalias{section}{appendix}
\crefname{section}{Appendix}{Appendices}
\Crefname{section}{Appendix}{Appendices}
\input{text/appendix/appendix}
\end{appendices}

\end{document}

%% file: our_imports.tex
\usepackage{enumitem}
\usepackage{xcolor}
\usepackage{tikz}
\usepackage{bbm}
\usepackage[page]{appendix}
\usepackage{wrapfig}
\usepackage{natbib} 
\usepackage{multirow}
\hypersetup{colorlinks=true, allcolors=red}
\usepackage{cleveref}
\AtBeginDocument{%
  \hypersetup{
    colorlinks=true,
    linkcolor=orange,
    citecolor=orange,
  }
}
\usepackage{array}
\usepackage{graphicx}
\usepackage{multirow}
\usepackage{booktabs}
\usepackage{makecell}
\usepackage{float}

\newcolumntype{P}[1]{>{\centering\arraybackslash}p{#1}}

\input{math_commands}

\usepackage{pifont}

\newcommand{\pce}{$\text{PCE}(u) \ $}

\theoremstyle{definition}
\newtheorem{definition}{Definition}

\definecolor{TabBlue}{HTML}{1F77B4}    
\definecolor{TabOrange}{HTML}{FF7F0E}  
\definecolor{TabGreen}{HTML}{2CA02C}   
\definecolor{TabRed}{HTML}{D62728}     
\definecolor{TabPurple}{HTML}{9467BD}  
\definecolor{TabBrown}{HTML}{8C564B}   
\definecolor{TabPink}{HTML}{E377C2}    
\definecolor{TabGray}{HTML}{7F7F7F}    
\definecolor{TabOlive}{HTML}{BCBD22}   
\definecolor{TabCyan}{HTML}{17BECF}    

\newcommand{\blackline}{{%
  \tikz[baseline=-0.5ex] \draw[very thick, color=black] (0,0) -- (0.3,0);%
}}
\newcommand{\reddashedline}{{%
  \tikz[baseline=-0.5ex] \draw[very thick, dashed, color=TabRed] (0,0) -- (0.3,0);%
}}
\newcommand{\graydashedline}{{%
  \tikz[baseline=-0.5ex] \draw[very thick, dashed, color=TabGray] (0,0) -- (0.3,0);%
}}

\newcommand{\orangeline}{{\tikz[baseline=-0.5ex] \draw[very thick, color=TabOrange] (0., 0.) -- (0.3, 0.);}}
\newcommand{\cyanline}{{\tikz[baseline=-0.5ex] \draw[very thick, color=TabCyan] (0., 0.) -- (0.3, 0.);}}
\newcommand{\grayline}{{\tikz[baseline=-0.5ex] \draw[very thick, color=TabGray] (0., 0.) -- (0.3, 0.);}}
\newcommand{\redline}{{\tikz[baseline=-0.5ex] \draw[very thick, color=TabRed] (0., 0.) -- (0.3, 0.);}}
\newcommand{\blueline}{{\tikz[baseline=-0.5ex] \draw[very thick, color=TabBlue] (0., 0.) -- (0.3, 0.);}}
\newcommand{\greenline}{{\tikz[baseline=-0.5ex] \draw[very thick, color=TabGreen] (0., 0.) -- (0.3, 0.);}}
\newcommand{\purpleline}{{\tikz[baseline=-0.5ex] \draw[very thick, color=TabPurple] (0., 0.) -- (0.3, 0.);}}

\crefname{section}{Section}{Sections}
\Crefname{section}{Section}{Sections}

\crefname{appendix}{Appendix}{Appendices}
\Crefname{appendix}{Appendix}{Appendices}

\newcommand{\appref}[1]{Appendix~\ref{#1}}

%% file: math_commands.tex
\usepackage{amsmath,amsfonts,bm}

\def\eqref#1{equation~\ref{#1}}

\def\1{\bm{1}}

\def\rs{{\textnormal{s}}}

\def\ve{{\bm{e}}}

\def\vh{{\bm{h}}}

\def\vp{{\bm{p}}}

\def\vv{{\bm{v}}}

\def\vx{{\bm{x}}}

\DeclareMathAlphabet{\mathsfit}{\encodingdefault}{\sfdefault}{m}{sl}
\SetMathAlphabet{\mathsfit}{bold}{\encodingdefault}{\sfdefault}{bx}{n}

\def\gD{{\mathcal{D}}}

\def\gH{{\mathcal{H}}}
\def\gI{{\mathcal{I}}}

\def\gR{{\mathcal{R}}}
\def\gS{{\mathcal{S}}}

\def\gU{{\mathcal{U}}}

\newcommand{\logits}{\mathrm{logits}}


%% file: text/abstract.tex
\begin{abstract}
  Popularity bias is a pervasive problem in recommender systems, where recommendations disproportionately favor popular items. This not only results in "rich-get-richer" dynamics and a homogenization of visible content, but can also lead to misalignment of recommendations with individual users’ preferences for popular or niche content. This work studies popularity bias through the lens of user-recommender alignment. To this end, we introduce \textit{Popularity Quantile Calibration}, a measurement framework that quantifies misalignment between a user’s historical popularity preference and the popularity of their recommendations. Building on this notion of popularity alignment, we propose \textit{SPREE}, an inference-time mitigation method for sequential recommenders based on activation steering. 
  SPREE identifies a popularity direction in representation space and adaptively steers model activations based on an estimate of each user's personal popularity bias, allowing both the direction and magnitude of steering to vary across users. 
  Unlike global debiasing approaches, SPREE explicitly targets alignment rather than uniformly reducing popularity. Experiments across multiple datasets show that SPREE consistently improves user-level popularity alignment while preserving recommendation quality. 
\end{abstract}

%% file: text/introduction.tex
\section{Introduction}

With the rapid expansion of online content, recommender systems have become curators and gatekeepers, deciding what we see and what we do not see. Most recommender systems disproportionally surface items that are already well-known and heavily exposed---an effect commonly referred to as \emph{popularity bias} \cite{abdollahpouri2017controlling, abdollahpouri2019managing, klimashevskaia2024survey}.
The resulting ``rich-get-richer'' dynamics cause problems across a range of domains. For example, on news and social media platforms, popularity bias can exacerbate the spread of populist perspectives and filter bubbles \citep{yalcin2022evaluating, ciampaglia2018algorithmic}. 

On most media streaming platforms, less popular, upcoming creators are dependent on recommender systems to gain traction. Creator popularity can mirror structural inequalities. For instance, female artists are historically underrepresented, and popularity-biased recommenders reinforce this inequality \cite{carnovalini2025popularity}. 
Since creators are typically compensated based on how much their work is consumed, this bias effectively reduces their livelihood. 
Further, the resulting lack of artist diversity
``can amplify dominant cultural norms, marginalizing underrepresented groups and their cultural expressions''  \cite{moradi2025embedding}. 
Thus, if popularity bias is not addressed, underrepresented creators are negatively impacted, both financially and culturally.

Popularity bias can also be detrimental to the user experience. 
Recommendations risk becoming bland or homogeneous, failing to reflect a user’s individual taste for popular (or niche) content \citep{abdollahpouri2020connection,abdollahpouri2021user,lichtenberglarge}. This undermines the core promise of personalization and is ultimately a threat to platform providers.
To gauge the impact on user experience, we study the recommender’s \emph{alignment} with human preferences, that is, whether the system captures an individual user’s preference for popular versus niche content. Measuring popularity bias from the user’s perspective has a practical benefit: While small content providers typically have little leverage to lobby for reducing popularity bias, platforms care about user experience. Enabling user-centric measurement and mitigation of popularity bias is of actionable economic interest to platforms, with positive side effects on creators and content diversity.

Yet measuring recommender-user popularity alignment can be challenging as user preferences can be multimodal and complex and we like to capture them as precisely as possible. 
To this end, we introduce \textit{popularity quantile calibration}, a new measurement framework that quantifies popularity bias as the calibration error between the popularity distribution of a user’s historical preferences and that of the model’s recommendations.

Building on this personalized notion of bias, we propose a mitigation method for transformer-based sequential recommenders. 
Our method--which we call \textit{SPREE}---adaptively steers predictions at inference time (i.e., without re-training) to align the recommender's popularity bias with the user's popularity preference.
We summarize our contributions as follows: 

\begin{itemize}
    \item We conduct an analysis of existing metrics for popularity bias through the lens of measurement theory (\Cref{sec:measurement_theory,sec:existing_metrics}) and provide desiderata for quantifying the discrepancy between recommenders and users (\Cref{sec:desiderata}). 
    \item We propose a new metric to measure alignment of recommender systems with user popularity preferences (\Cref{sec:calibration,sec:diagnostic}).
    \item We propose a simple, lightweight mitigation method that reduces misalignment of transformer-based sequential recommenders. The method does not require re-training and mitigates bias individually per user (\Cref{sec:mitigating}).   
    \item Experimental evaluation of our approach on a range of datasets across different domains demonstrates that it provides the most effective alignment-performance trade-off among baseline mitigation methods (\Cref{sec:experiments}).
\end{itemize}

\noindent 

%% file: text/related-work.tex
\section{Related Work}
Our work builds on two bodies of literature: measuring popularity bias in recommender systems, and methods to mitigate it. We discuss measurement approaches in depth in \Cref{sec:existing_metrics} through the lens of measurement theory. Here we survey mitigation strategies and activation steering, the technical toolkit underlying our mitigation method, SPREE.

\paragraph{Mitigating popularity bias}
Mitigating strategies can be grouped into two line of works: training-time methods and inference-time methods.
The first line of work mitigates popularity bias either by
modifying the training data distribution or the training objective.
\citet{boratto2021combining} combine interaction sampling that balances
head–tail training pairs with a regularization term that reduces correlation
between predicted relevance and item popularity.
\citet{abdollahpouri2017controlling} similarly control head–tail exposure
through a regularization-based learning-to-rank formulation.
\citet{ning2024debiasing} introduce a personal-popularity-aware counterfactual
framework that requires learning auxiliary popularity estimators during
training, while exposing tunable popularity control at inference. Some solutions
modify the item embedding representations: \citet{xv2022neutralizing} add
regularizers to learn popularity-neutral item embeddings, while
\citet{ren2022mitigating} analyze popularity bias from a gradient perspective
and propose a mitigation that relies on training-time gradient statistics with a
subsequent item embedding adjustment stage. Finally, several works operate at
the sampling level, showing that negative sampling choices strongly affect
head-tail outcomes \cite{prakash2024evaluating}, and proposing
popularity-corrected negative sampling within contrastive/pairwise objectives to
counteract skewed training data \cite{lu2025popularity}. Closer to our method, the
second line of work intervene at inference time, without requiring
retraining or modifying the optimization objective. Early work by
\citet{zhang2010niche} promotes niche items in top-$N$ recommendation via
popularity-aware re-scoring. \citet{abdollahpouri2019managing} propose
personalized post-processing re-ranking to manage popularity bias without
changing the base model. In~\cite{abdollahpouri2021user}, \emph{Calibrated
Popularity} (CP) is introduced that re-ranks the items balancing relevance and a
user-specific popularity calibration term. 
Closest to our work are methods that directly manipulate internal signals at inference: 
\citet{abbattista2024enhancing} use personalized popularity information in sequential recommendation to balance relevance scores and user-item popularity scores, and \citet{mallamaci2025balancing} extend the idea of using personalized popularity at sub-item level.


\paragraph{Activation steering}
Activation steering intervenes on a model’s internal representations at inference time by adding or subtracting direction vectors in activation space to bias outputs toward or away from an encoded concept, without retraining. It has been widely used to align large language model behavior \citep{liuiterative,cao2024personalized,qiu2024spectral} for bias reduction \citep{adila2024discovering,sharma2025optimal} among others. The steering strength $\lambda$ may be global \citep{liu2023context,arditi2024refusal}, scaled by activations \citep{stolfo2024improving}, or -- closest to our approach -- estimated from the input \citep{hedstrom2025steer}. The most closely related work is PopSteer \citep{ahmadov2025opening}, which also applies activation steering in sequential recommenders to address popularity bias. Our approach differs in four key aspects: (i) we align recommendations to user-specific popularity preferences (S3), whereas PopSteer targets global popularity effects (S2); (ii) PopSteer therefore does not support input-dependent steering magnitudes; (iii) PopSteer steers in a sparse autoencoder (SAE) latent space, while we intervene directly in the model’s activation space omitting reconstruction errors of the SAE; and (iv) PopSteer intervenes at the user embedding $(x_L)$, whereas SPREE automatically selects the best block $\ell$, allowing to impact the residual stream early on.

%% file: text/popularity_bias_problem.tex
\section{The Problem of Popularity Bias}
Popularity bias — the tendency of recommender systems to disproportionately surface already well-known items — creates measurable harms for users, creators, and platforms alike. We first characterize these harms across stakeholders, then formalize the problem setting that grounds our measurement and mitigation framework.

\subsection{Harms of Popularity Bias across Stakeholders}
\label{sec:stakeholders}
Recommender systems operate as multi-sided platforms, simultaneously serving users seeking relevant content, creators seeking audience, and platforms seeking engagement — while shaping broader cultural discourse. Popularity bias impacts this ecosystem, creating harms across all stakeholders \citep{abdollahpouri2019multi}.

\textit{Content creators} of non-mainstream content are systematically denied visibility, causing (i) economic harm through reduced streams, clicks, and revenue, and (ii) representational harm as their voices and impact are suppressed \citep{deldjoo2024fairness}. Critically, popularity bias (iii) reinforces structural inequalities such as the chronic underrepresentation of female artists \citep{moradi2024advancing} and (iv) incentivizes self-censorship as creators may optimize for what algorithms reward \citep{raffa2021algorithmic}.

\textit{Users} experience popularity bias as (i) undermined user fairness \citep{ekstrand2022fairness}, with service inequality documented based on users' preferences \citep{yalcin2022evaluating}, language \citep{elahi2021investigating}, gender and age \citep{ekstrand2018all,lesota2021analyzing}. It further (ii) drives content homogenization and filter bubbles \citep{ciampaglia2018algorithmic}, risking erosion of user preferences, self-image, and mental wellbeing \citep{smith2024instagram}.

\textit{Platforms} face popularity bias through (i) threatened long-term user satisfaction and retention, (ii) content creator churn as niche providers leave platforms, and (iii) exposure to legal liability under algorithmic accountability frameworks \citep{EUAIAct,DSA}.

\textit{Society} as a whole is also affected as popularity bias (i) amplifies dominant cultural norms while marginalizing minority expressions \citep{moradi2025embedding}. As a consequence, (ii) affected groups' sense of belonging may erode through amplified misrepresentation \citep{blodgett2020language,kay2015unequal}. Platform-specific harms include (iii) political polarization \citep{shmargad2020sorting} and (iv) gate-keeping artistic legacy and cultural heritage \citep{porcaro2021diversity}.

\subsection{Formal Problem Setting}
\label{sec:problem_setting}
Having established the societal stakes of popularity bias, we now formally define the problem setting for measuring and mitigating it in recommender systems. Let $\gU = \{u_1, \dots, u_{|\gU|}\}$ denote the set of users and $\gI = \{i_1, \dots, i_{|\gI|}\}$ the set of items of a recommender system. 
We consider the problem of generating a personalized recommendation list $\gR_u=[i_1, \dots, i_K]$ of $K$ relevant items based on the user's past interaction history $\gH_u = [i_1, i_2, \dots, i_{T_u}],$
where $T_u$ is the number of interactions user $u$ had in the past. 
$\gS=\bigcup_{u\in \gU} \gR_u$ denotes the set of recommended items for all users. 
A system suffers from popularity bias if popular items are over-represented in the system's recommendation lists $\gR_u$. We will discuss in more detail what ``over-represented'' can mean in \Cref{sec:measuring}. The popularity $s(i)$ of an item $i$ is defined as $s(i) = \sum_{u \in \gU}\sum_{j=1}^{T_u} \mathbbm{1}_{\gH_u(j) = i}$, where $\gH_u(j)$ is the $j$-th element in $\gH_u$, and represents the total number of times item $i$ has been interacted with by all users $\gU$. Analogously, $\hat{s}(i)= \sum_{u \in \gU}\sum_{j=1}^{K} \mathbbm{1}_{\gR_u(j) = i}$, where $\gR_u(j)$ is the $j$-th element in $\gR_u$, denotes the total number of times item $i$ has been recommended by the system. In other words, $s$ is a function of the histories $\gH_u$, $u \in \gU$, and $\hat{s}$ a function of the recommendations $\gS$. 

In this work, we view a user's popularity preference as a conditional distribution over $s$, and denote it by $p(s | u)$.
Users with niche tastes put more probability mass on low $s$, while those with more mainstream tastes favor high $s$.
The true preference is unknown but can be estimated empirically from 
the user’s past interactions.
Similarly, a recommender system induces a distribution over $s$ (conditioned on $u$) based on how likely it is to recommend popular content, and we denote this by $q(s | u)$.
An \emph{aligned} recommender should reproduce the user-specific distribution, such that 
$q(s | u)$ closely matches $p(s | u)$.
We go into detail on how to estimate and compare these distributions in \Cref{sec:calibration}. Before we propose our measuring framework, we next review existing metrics and assess which understanding of popularity bias they aim to measure.

%% file: text/measuring-bias-past-work.tex
\section{Measuring Popularity Bias}
\label{sec:measuring}
Popularity bias has been a persistent problem in the literature on recommender systems \citep{klimashevskaia2024survey}, motivating a large body of work on metrics to quantify the popularity bias and methods to mitigate it. However, popularity bias is an abstract and contested concept: different threads of the literature implicitly adopt different theoretical understandings of what constitutes “bias,” and corresponding measurement instruments are often used for related but distinct concepts such as item fairness, diversity, novelty, or serendipity \citep{klimashevskaia2024survey}. As a result, existing metrics are frequently compared or optimized without a clear account of which underlying notion of popularity bias they operationalize.

In this section, we use \textit{measurement theory} as a guiding framework to summarize and assess how popularity bias has been measured in prior work. Specifically, we (i) disentangle competing systematized concepts of popularity bias that arise from different interpretations (\Cref{sec:existing_metrics}); (ii) analyze common measurement instruments of user-centric popularity bias through established validity lenses (\Cref{sec:desiderata}); and (iii) derive  desiderata that motivate a new measurement framework for popularity bias (\Cref{sec:calibration,sec:diagnostic}). While the primary goal of the paper is not to provide a full measurement-theoretic analysis of popularity bias, we use the well-defined terminology of measurement theory to guide our review of existing metrics and to crystallize distinctions between prior metrics and our proposed measurement framework.

\subsection{A Measurement Theory Perspective}
\label{sec:measurement_theory}
Following \citep{wallach2025position,chouldechova2024shared}, we adopt the measurement theory framework of \citet{adcock2001measurement}. The framework distinguishes four stages: a \textit{background concept}, capturing the range of meanings and interpretations associated with a phenomenon; a \textit{systematized concept}, which provides a precise and explicit definition of what is to be measured; a \textit{measurement instrument}, which operationalizes the systematized concept; and the resulting \textit{measurements}. 

These stages are connected through four processes \citep{adcock2001measurement}: \textit{systematization}, which narrows the background concept into a systematized concept; \textit{operationalization}, which implements the systematized concept in a measurement instrument; \textit{application}, which uses the instrument to obtain measurements; and, critically, \textit{interrogation}, a feedback process that evaluates whether each stage preserves validity with respect to the preceding ones. For the purposes of this work, we neglect the application stage and first focus on the systematization and operationalization stage in the next section. This serves to clearly distinguish the understanding of popularity bias we adopt in this work from related but distinct interpretations.

\subsection{Systematization and Operationalization: Overview of Existing Measurement Approaches}
\label{sec:existing_metrics}
\begin{sloppypar}
Popularity bias is a contested background concept in the recommender-systems literature, with multiple theoretical understandings resulting in distinct systematized concepts. Each of these systematized concepts has, in turn, been operationalized through a variety of measurement instruments. Please refer to \citep{klimashevskaia2024survey} for a detailed survey.
\end{sloppypar}

We highlight three strains of understanding in the literature. \Cref{tab:existing_metrics} summarizes their systematized concepts of popularity bias and the representative measurement instruments associated with each.
\textit{(S1) the absolute popularity of the recommended set of items $\gS$} is interested in a global popularity measure of the system. Metrics operationalizing (S1) are average recommendation popularity \citep{abdollahpouri2019managing} and its log variant \citep{lichtenberglarge}.
\textit{(S2) Concentration of recommendation frequency} measures exposure differences across items. Popular metrics comprise the share of items recommended, i.e., coverage \citep{adomavicius2011improving}, Shannon Entropy of the recommendation distribution \citep{carnovalini2025popularity}, and the Gini- \citep{gini1921measurement} and Herfindahl-Index \cite{deldjoo2024understanding}.
\textit{(S3) The difference in popularity between interaction history and recommendation for a given user} quantifies user–recommender alignment. Metrics include popularity lift (PL) \citep{abdollahpouri2020connection}, which is the average, normalized distance between recommended and consumed items for a given user $u$. LogPopLift \citep{lichtenberglarge} drops the normalization and reports average log popularity. User popularity deviation (UPD) partitions the popularity score $s$ into disjoint bins (low, medium, and high popularity) and then computes the Jensen-Shannon divergence (JSD) between the induced empirical distributions over these bins for the recommendation set $\gR_u$ and the user history $\gH_u$.

We note that operationalization choices of (S3) vary substantially in how item popularity $s(i)$ is transformed—e.g., using the identity (PL), logarithmic scaling (LogPopLift), or discretization into bins (UPD)—but also in how user histories and recommendation lists are compared—notably via relative differences (PL), absolute differences (LogPopLift), or distributional distances (UPD). This raises the question of how to make such operationalization choices and their consequences for the metrics' validity. To be transparent about these choices, we derive desiderata for validity-preserving  operationalization and show that existing metrics fail to meet them in the following section. Since our ultimate goal is user-recommender  popularity alignment, we henceforth focus on category (S3).

\input{tables/concept} 

\subsection{Interrogation: Desiderata for Measurement Instruments}
\label{sec:desiderata}
We focus on three interrogation lenses \citep{jacobs2021measurement,wallach2025position}: \textit{content validity}, \textit{external validity}, 
and \textit{construct reliability}. Using each lens to analyze existing measurement instruments, we identify limitations in their validity and derive desiderata to circumvent these limitations. \Cref{tab:desiderata} summarizes which desiderata are satisfied by each metric.

\textit{Content validity} refers to the extent to which a measurement instrument represents all facets of a systematized concept.
In the context of \textit{(S3) difference in popularity between $\gH_u$ and $\gR_u$}, we identify three facets of "difference in popularity" that we argue should be captured: \textit{(D1) ordinality} (is the popularity of $\gR_u$ larger than that of $\gH_u$ or vice versa?); \textit{(D2) cardinality} (are small deviations between $\gH_u$ and $\gR_u$ captured by the measurement instrument?); \textit{(D3) dispersion} (are item popularities in $\gR_u$ more concentrated than in $\gH_u$?). Capturing dispersion is crucial for matching more complex popularity preferences, as a user may, for example, prefer both highly popular and very niche items. 
Both LogPopDiff \citep{lichtenberglarge} and PL \citep{abdollahpouri2020connection} satisfy (D1) and (D2), but not (D3) because they rely only on the popularity means for $\gR_u$ and $\gH_u$.
UPD \citep{abdollahpouri2021user} fails to satisfy all three desiderata, which we illustrate via counterexamples in \Cref{fig:content_validity}. The first two plots show when UPD fails to satisfy (D1) and (D2), using hypothetical histories and recommendations for a pair of users, $u$ and $u'$. Recall that UPD discretizes the popularity score $s$ into several bins (low, medium, and high; indicated by the red dotted thresholds, \reddashedline), creating discrete distributions which can be compared via the Jensen-Shannon divergence. Since the Jensen-Shannon divergence is symmetric, UPD obtains the same popularity bias for these users, even though $\gR_u$ (\grayline) recommends overly popular items and $\gR_{u'}$ (\graydashedline) overly niche items, violating (D1). Further, UPD does not distinguish items within each bin, so it can capture neither cardinality (\Cref{fig:content_validity}, middle) nor dispersion (\Cref{fig:content_validity}, right) sufficiently, hence violating (D2) and (D3).

\input{figures/content_validity}

\textit{External validity} refers to the extent to which a measurement instrument supports stable comparisons across contexts beyond those in which it was originally developed or validated. In recommender systems, such comparisons across contexts can  be different item bases $\gI$, user bases $\gU$, or different temporal snapshots of the same user and item base. 
To enable such comparisons, the measurement instrument must be \textit{(D4) scale-invariant}, i.e. the size of users, items, or interactions should not affect the bias metric. While (D4) is satisfied by PL and UPD, LogPopDiff violates it by computing absolute differences in popularity scores between $\gH_u$ and $\gR_u$. Because absolute popularity depends on the total number of interactions, its scale varies across datasets and increases over time as user interactions accumulate.

\textit{Construct reliability} concerns the stability and consistency of measurements under perturbations. In recommender systems, user histories and recommendation lists are often small, making measurements particularly sensitive to sampling variability. A reliable measurement instrument should therefore yield robust values under small perturbations, such as the addition or removal of a single highly popular item. This motivates \textit{(D5) robustness to outliers}. By using mean aggregation, PL (and, to a lesser extent, LogPopDiff due to the log transform) is sensitive to extreme popularity values.

\input{tables/desiderata}

%% file: tables/concept.tex
\begin{table}[t!]
    \centering
    \caption{Overview of systematized concepts and corresponding measurement instruments for the background concept of popularity bias. Popularity bias has been understood at the system level (S1), as a function of all recommendations $\gS$; at the item level (S2), as a function of the item set $\gI$ and the recommendations $\gS$; and at the user level (S3), as a function of the user $u$.}

    \scalebox{0.8}{
    \begin{tabular}{P{3.5cm}|c}
        \toprule
        Systematized Concept & Measurement Instrument \\
        \midrule
        \multirow{4}{=}{\parbox{3.2cm}{\centering
        (S1) Absolute popularity of the recommended set of items $\gS$ }}
        & \makecell[c]{Average Recommendation Popularity \citep{abdollahpouri2019managing} \\[4pt]
        $\displaystyle 
        \text{ARP}(\gS) = \frac{\sum_{i \in \gS}s(i)}{|\gS|}$}  \\
        \cmidrule(l){2-2}
        & \makecell[c]{Average Log Recommendation Popularity \citep{lichtenberglarge}\\
        $\displaystyle 
        \text{ALRP}(\gS) = 
        \frac{\sum_{i \in \gS}\log s(i)}{|\gS|}
        $} \\
        \cmidrule(l){1-2}
        
        \multirow{14}{=}{\parbox{3.2cm}{\centering
            (S2) Concentration of recommendation frequency\\
            on some items in catalogue $\gI$}}
        & \makecell[c]{Coverage or Aggregate Diversity \citep{adomavicius2011improving} \\[4pt]
        $\displaystyle 
        \text{Coverage}(\gS, \gI) = \frac{|\gS|}{|\gI|}$} \\
        \cmidrule(l){2-2}
        
        & \makecell[c]{Shannon Entropy \citep{carnovalini2025popularity} \\ 
        $\displaystyle \text{H}(\gS, \gI) = -\sum_{i \in \mathcal{I}} \phi(i)\log \phi(i)$, \\ where $\phi(i)=\frac{\hat{s}(i)}{\sum_{j \in \gI} \hat{s}(j)}$} \\
        
        \cmidrule(l){2-2}
        
        & \makecell[c]{Herfindahl Index \citep{deldjoo2024understanding} \\[4pt]
        $\displaystyle 
        \text{HHI}(\gS, \gI) = 
        \sum_{i \in \gI} 
        \left(
        \frac{\hat{s}(i)}{\sum_{j \in \gI} \hat{s}(j)}
        \right)^2$} \\  
        \cmidrule(l){2-2}
        
        & \makecell[c]{Gini Index \citep{gini1921measurement, zeng2012reinforcing, braun2023metrics} \\[4pt]
        $\displaystyle 
        \text{Gini}(\gS, \gI) = 
        \frac{\sum_{i \in \gI}(2i-|\gI|-1)\hat{s}(i)}
        {|\gI|\sum_{i \in \gI} \hat{s}(i)}$,\\
        with $\hat{s}(i)$ ordered in ascending order} \\
        \cmidrule(l){1-2}

        \multirow{8}{=}{\parbox{3.2cm}{\centering
            (S3) Difference in popularity between interaction history $\gH_u$\\
            and recommendations $\gR_u$ for a given user $u$}}
        & \makecell[c]{Popularity Lift \citep{abdollahpouri2020connection} \\[4pt]
        $\displaystyle 
        \text{PL}(u) = 
        \Big(
        \frac{\sum_{i \in \gR_u}s(i)}{|\gR_u|} -
        \frac{\sum_{i \in \gH_u}s(i)}{|\gH_u|}
        \Big) \bigg/ 
        \frac{\sum_{i \in \gH_u}s(i)}{|\gH_u|}$} \\
        \cmidrule(l){2-2}

        & \makecell[c]{Log Popularity Difference \citep{lichtenberglarge} \\[4pt]
        $\displaystyle 
        \text{LogPopDiff}(u) =
        \frac{\sum_{i \in \gR_u}\log s(i)}{|\gR_u|} -
        \frac{\sum_{i \in \gH_u}\log s(i)}{|\gH_u|}$} \\
        \cmidrule(l){2-2}

        & \makecell[c]{User Popularity Deviation \citep{abdollahpouri2021user} \\[4pt]
        $\displaystyle 
        \text{UPD}(u) = \text{JSD} \!\big(p(s | u), q(s | u)\big)$} \\
        
        \bottomrule
    \end{tabular}
    }
    \label{tab:existing_metrics}
\end{table}

%% file: figures/content_validity.tex
\begin{figure}
    \centering
    \vspace{-0pt}
    \includegraphics[width=\linewidth]{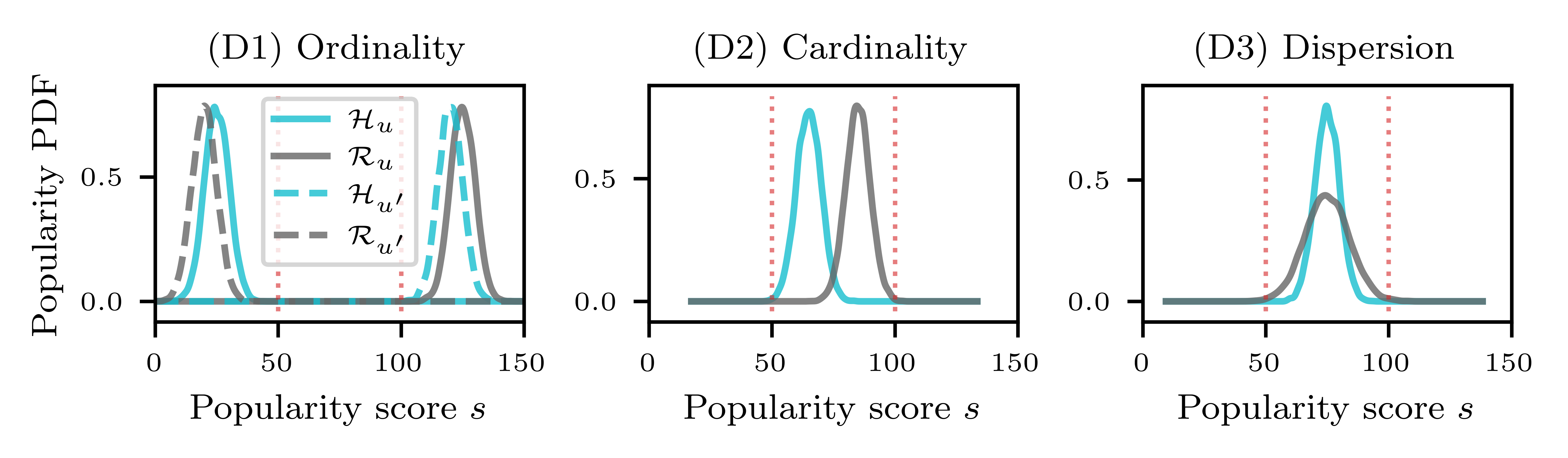}
    \caption{Problematic cases for the UPD metric regarding content validity: \textbf{Left:} Recommendations for users $u$ (---) and $u'$ (- -) are positively and negatively biased, respectively. Nevertheless, UPD assigns the same popularity bias due to the symmetry of the underlying Jensen–Shannon divergence. \textbf{Middle:} UPD does not measure popularity bias within bins (red dashed lines). \textbf{Right:} Since UPD does not differentiate within bins, no popularity bias is reported despite different popularity variance in $\gH_u$ and $\gR_u$.}
    \label{fig:content_validity}
    \vspace{-1em}
\end{figure}

%% file: tables/desiderata.tex
\begin{table}[t]
\centering
\caption{Comparison of popularity bias metrics that operationalize (S3), the popularity alignment between the user and the recommender, with respect to our desiderata.}
\scalebox{0.77}{
\begin{tabular}{l|ccccc}
\toprule
Desideratum 
 & \makecell{PopLift \\ \citep{abdollahpouri2020connection}}
 & \makecell{LogPopDiff \\ \citep{lichtenberglarge}}
 & \makecell{UPD \\ \citep{abdollahpouri2021user}}
 & \makecell{PCE \\ (ours)}
 & \makecell{Calib. Curve \\ (ours)} \\
\midrule
(D1) Ordinality             & \checkmark & \checkmark & $\times$   & $\times$ & \checkmark \\
(D2) Cardinality             & \checkmark & \checkmark & $\times$   & \checkmark & \checkmark\\
(D3) Dispersion     & $\times$   & $\times$   & $\times$ & \checkmark & \checkmark \\
(D4) Scale-invariance         & \checkmark & $\times$   & \checkmark & \checkmark & \checkmark\\
(D5) Robustness to outliers        & $\times$   & $\times$   & \checkmark & \checkmark & \checkmark\\
\bottomrule
\end{tabular}
}
\label{tab:desiderata}
\end{table}

%% file: text/measuring-bias-ours.tex
\subsection{Alignment as Calibration}
\label{sec:calibration}

Equipped with the desiderata in \Cref{tab:desiderata}, our goal is to design a measurement framework that captures differences in direction (D1), magnitude (D2) and variance (D3), while maintaining favorable mathematical (D4) and statistical properties (D5). Since we aim to account for higher-order moments (D3), we consider the full distributions, rather than relying on simple summary statistics such as means. Assessing whether a predictive distribution, $q(s | u)$, approximates the ground-truth empirical distribution, $p(s | u)$, is the central problem of calibration. We build on the rich literature on calibration \citep{wang2023calibration,kuleshov2018accurate,guo2017calibration} for our measurement framework.
In contrast to prior work on calibration in recommender systems---which focuses on categorical outcomes such as genres \citep{steck2018calibrated} or binned popularity~\citep{abdollahpouri2021user}---our definition operates on the inherently numerical scale of $s$. This preserves the cardinality of $s$ (D2) without resorting to discretization (like UPD), thereby avoiding its associated pitfalls (see \Cref{fig:content_validity}). Inspired by calibration in regression \citep{kuleshov2018accurate, levi2022evaluating, song2019distribution}, we compare $p(s | u)$ and $q(s | u)$ via their quantile functions, evaluating their overlap at a finite set of quantile levels.

More formally, let $F_q$ denote the \emph{cumulative distribution function} (CDF) of $q(s | u)$, 
$$
F_q(s) = \mathbb{P}_{x \sim q(\rs | u)}(x \le s) \quad \text{and} \quad
F_q^{-1}(\tau) = \inf \{ s : \tau \le F_q(s) \}
$$
where $F_q^{-1}$ is the corresponding quantile function.
Intuitively, the CDF $F_q(s)$ measures the fraction of items with a popularity score smaller than or equal to $s$, while the quantile function $F_q^{-1}(\tau)$ gives the popularity value below which a fraction $\tau$ of the items fall.

\input{figures/metric}

\begin{definition}[Popularity quantile-calibrated]
A recommender is quantile-calibrated in popularity for user $u$ if
$$
\mathbb{P}_{x \sim p(s | u)}(x \le F_q^{-1}(\tau)) = \tau \quad \text{for all } \tau \in [0,1].
$$
\label{def:calibration}
\end{definition}
Intuitively, this means that the share of items in the user’s history with popularity below the $\tau$-quantile of the recommendation distribution equals $\tau$ itself. For example, if $\tau = 0.8$ and the 80th percentile of the  recommender's distribution for user $u$, $p(s |u)$, corresponds to a popularity score of $s = 500$ (i.e. $F_q^{-1}(0.8) = 500$), then $80\%$ of the items in the user’s history $\mathcal{H}_u$ should also have popularity scores below $500$. In other words, if the CDFs of the user’s interaction history and the recommender’s output distribution match, then the recommender reproduces the user’s individual popularity preference. In practice, we select a finite set of $m$ quantile levels for $\tau$ at which to compare the CDFs. In doing so, we implicitly perform a form of binning; however, the binning is applied to the user's recommender distribution $q(s | u)$ rather than\ the global popularity values $p(s)$ (as in UPD). Consequently, binning is performed individually for each user, ensuring that recommended items are necessarily spread across $m$ bins and can never all collapse into a single bin.

\subsection{Proposed Measurement Instruments}
\label{sec:diagnostic}

We propose two measurement instruments that operationalize (S3) based on the quantile-calibration notion in \Cref{def:calibration}: 

\paragraph{Popularity Calibration Curve.}
Analogous to reliability diagrams in regression \citep{kuleshov2018accurate}, we use a calibration curve that depicts the miscalibration between $p(s | u)$ and $q(s | u)$ for different levels of $\tau \in [0,1]$.
For each user $u$, we consider a set of $m$ quantile levels $0 \le \tau_1 < \tau_2 < \dots < \tau_m \le 1$.
At each level $\tau_j$, we compute the corresponding quantile threshold $s^{(\tau_j)} = F_q^{-1}(\tau_j)$ from the recommendation distribution $q(s | u)$ (i.e., the popularity $s$ below which a share of $\tau_j$ items in $\gR_u$ fall). Then, we estimate the empirical fraction of items in the user’s history that are less popular than $s^{(\tau_j)}$:
\begin{equation} \hat{\tau}_j = 
\frac{1}{|\gH_u|} \sum_{i \in \gH_u} \mathbbm{1}\{ s(i) \le s^{(\tau_j)} \}. 
\label{eq:tau_hat}
\end{equation}

Plotting the pairs ${(\tau_j, \hat{\tau}_j)}_{j=1}^{m}$ yields a \emph{popularity calibration curve}, an example of which is shown in \Cref{fig:popularity_calibration_curve} (right).
A perfectly calibrated recommender corresponds to the diagonal $\tau_j = \hat{\tau}_j$. When a recommender generates overly popular items, the calibration curve lies above the identity line. When a recommender predicts overly niche items, the calibration curve lies below the identity line. Thus, the calibration curve satisfies (D1) ordinality. Cardinality (D2) is controlled by the number of quantile levels \(m\): a higher \(m\) results in more fine-grained comparisons of \(p(s|u)\) and \(q(s|u)\).
It can also represent (D3) dispersion: a curve above the identity line for low quantiles and below the identity line for high quantiles, as depicted in \Cref{fig:popularity_calibration_curve} (right), corresponds to a recommender with too small a variance (see the corresponding PDF in \Cref{fig:popularity_calibration_curve}, left); a reversed curve reflects too high a variance of \(q(s | u)\). Because the measure is defined in terms of quantiles, it depends only on the relative ordering of items with respect to \(s\). This makes it scale-invariant (D4) and, at the same time, robust to outliers (D5), since extreme values affect magnitudes but not ranks  — the latter being what our framework relies on.

\begin{sloppypar}
\paragraph{Popularity Calibration Error (PCE)} To obtain a numerical instrument, we aggregate deviations between $\tau$ and $\hat{\tau}$ over the $m$ quantile levels. The resulting \emph{popularity calibration error} is the analogue of the calibration error for regression~\citep{kuleshov2018accurate},
\begin{equation*}
\text{PCE}(u) = \frac{1}{m}\sum_{j=1}^{m} (\tau_j - \hat{\tau}_j)^2. 
\end{equation*}
\Cref{fig:popularity_calibration_curve} (right) illustrates PCE in the calibration curve. It is the average of the squared differences (\orangeline) between $\tau$ (here $\tau \in \{0, 0.2, 0.4, 0.6, 0.8, 1\}$) on the x-axis and $\hat{\tau}$ on the y-axis. It can also be understood as the differences in CDFs evaluated at the popularity levels $s^{(\tau_j)}$ induced by $\tau_j$ (\Cref{fig:popularity_calibration_curve}, middle). \pce ranges between $0$ and $1$, where $0$ corresponds to perfect calibration at the given set of $\tau_j$. Like the calibration curve, it satisfies (D2)-(D5). However, it sacrifices (D1) ordinality, since the squared difference loses the direction of the bias. This is a deliberate choice, as ultimately, we are interested in a global metric averaged over all users, where 0 indeed corresponds to zero bias (and not equally positive and negative bias). We obtain a global measure by averaging over all users,
$ \text{PCE} = \frac{1}{|\gU|} \sum_{u \in \gU} \text{PCE}(u). $
PCE evaluated at the top $K$ predictions is denoted by $\text{PCE}@K$. 
\end{sloppypar}

%% file: figures/metric.tex
\begin{figure*}[h!]
    \centering
    \includegraphics[width=\linewidth]{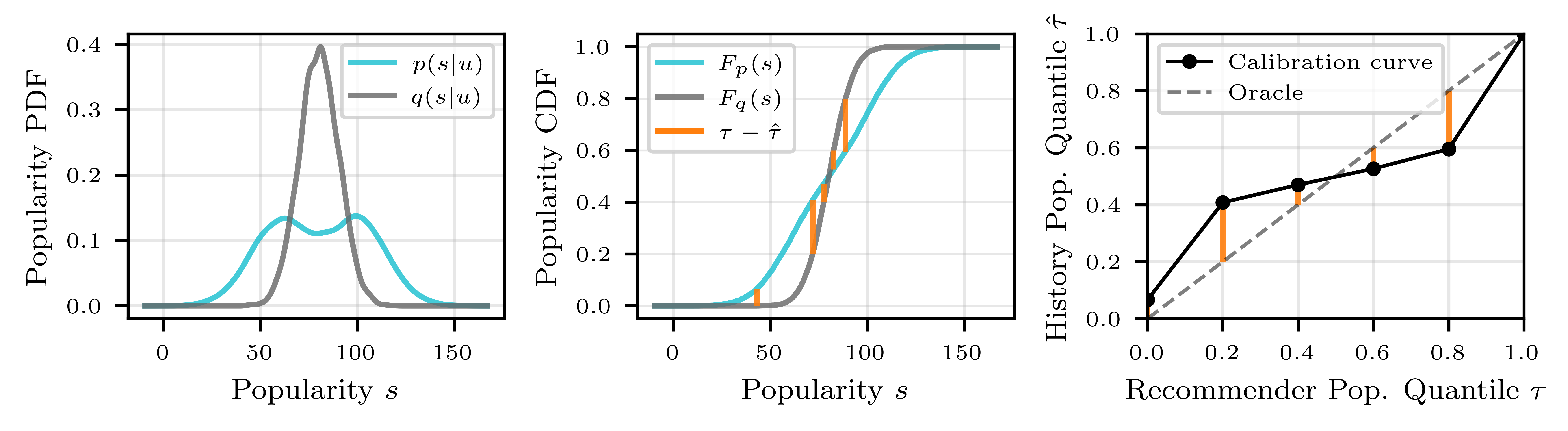}
    \caption{Visualization of \textit{popularity quantile calibration}. \textbf{Left:} In the toy example, the user’s popularity preference $p(s | u)$ is more widely spread than that of the recommender, $q(s | u)$. \textbf{Middle:} Consequently, their CDFs diverge. PCE measures the difference between the user CDF $F_p(s)$ and the recommender CDF $F_q(s)$ at fixed quantiles $\tau \in \{0, 0.2, 0.4, 0.6, 0.8, 1\}$ (orange line). \textbf{Right:} The resulting calibration curve deviates from the perfectly calibrated oracle diagonal indicating the extent of miscalibration.}

    \label{fig:popularity_calibration_curve}
\end{figure*}

%% file: text/mitigating-bias.tex
\section{Mitigating Popularity Misalignment}
\label{sec:mitigating}
Equipped with a framework for measuring popularity alignment, we now turn to our activation-steering method for mitigating popularity bias as defined in (S3).
Activation steering has been employed in language models and builds on the \textit{linear representation hypothesis} \citep{park2023linear}, which posits that high-level concepts are encoded approximately linearly in a model’s hidden states. The idea of activation steering is simple: by identifying a concept vector and adding or subtracting it from the internal representations at inference time, we can steer the model toward or away from specific behaviors associated with the concept. Recent work has shown that activation steering can guide LLMs toward more truthful \citep{li2023inference,li2025hidden}, instruction-following \citep{stolfo2024improving}, or safe responses \citep{liu2023context,hedstrom2025steer}, among others \citep{arditi2024refusal}. Its appeal lies in effectively controlling high-level concepts through a simple, post-hoc modification that requires no training of the base model.
We first provide background on how sequential recommenders process user sequences to predict relevant next items in \Cref{sec:transformers}. In \Cref{sec:steering}, we describe how we identify a popularity direction in a recommender’s activation space. \Cref{sec:steering_strength} then describes how this direction is used to reduce each user’s individual popularity misalignment.

\subsection{Background: Transformer-based Recommenders}
\label{sec:transformers}
Transformer-based sequential recommenders \citep{kang2018self,petrov2023gsasrec,li2020time,lin2020fissa,petrov2023generative,wu2020sse} such as SASRec \citep{kang2018self} are state-of-the-art models deployed on various online platforms \citep{pancha2022pinnerformer,zhai2024actions,Palumbo2025Text2TracksPM,Penha2025SemanticIF}.
They model a user’s interaction history as an ordered sequence of item embeddings and use self-attention to capture temporal preference dynamics.  
Given a sequence of user interactions $\gH_u = [i_1, i_2, \dots, i_{T}]$, the model predicts the next item to be interacted with and outputs a list of the $K$ most relevant items $\gR_u \subseteq \mathcal{I}$.
Consider a sequential recommender composed of $L$ stacked transformer blocks.  
Let $\vx_{t, \ell} \in \mathbb{R}^d$ denote the residual stream activation  of the $t$-th sequence position after block $\ell$, where $\ell \in \{0, \dots, L\}$.  
The input to the first block is the sum of learned item and positional embeddings,
\[
\vx_{t, 0} = \ve_{i_t} + \vp_t,
\]
where $\ve_{i_t}, \vp_t \in \mathbb{R}^d$ are the item and positional embedding vectors of the item at position $t$, respectively and $d$ is the dimensionality of the representation space.  
Each subsequent block $\ell$ transforms the sequence $\vx_{\ell-1} = [\vx_{1, \ell-1}, \dots, \vx_{T, \ell-1}]$ through a multi-head attention, feed-forward and layer-normalization sublayer,
$$
\begin{aligned}
\tilde{\vx}_{t, \ell} &= \tilde{\text{LN}}_\ell\Big(\vx_{t, \ell-1} + \big[\text{Attn}_{\ell}\big(\vx_{\ell-1}\big)\big]_t\Big), \\
\vx_{t, \ell} &= \text{LN}_\ell\Big(\tilde{\vx}_{t, \ell} + \text{MLP}_{\ell}\big(\tilde{\vx}_{t, \ell}\big)\Big).
\end{aligned}
$$
Here, $\text{Attn}_{\ell}(\cdot)$ denotes the causal multi-head self-attention layer, $[\cdot]_t$ indexes its $t$-th position, $\text{MLP}_{\ell}(\cdot)$ denotes the position-wise feed-forward network, $\tilde{\text{LN}}_{\ell}(\cdot)$  and $\text{LN}_{\ell}(\cdot)$ the layer normalization layers of block $\ell$. The residual stream activation $\vx_{t, \ell}$ is then passed through the next transformer block $\ell +1$. After the final block $L$, the representation of the last position $\vx_{T, L}$ serves as the \emph{user embedding} $\vh$.  Predictions for user $u$ are obtained by scoring all candidate items $i \in \gI$ via the dot product between the user embedding $\vh$ and item embeddings $\ve_i$, $\logits_{i} = \vh^\top \ve_i$. The top-$K$ items with the highest logit score for user $u$ form the recommendation set $\gR_u$,

\subsection{Controlling Popularity via Activation Steering}
\label{sec:steering}

To elicit the steering vector and the optimal steering positions \(t, \ell\) in the model, we generate two contrastive sets of artificial user sequences, \(\gD^+\) and \(\gD^-\), containing high- and low-popularity sequences, respectively. For this, we split the item set into disjoint subsets based on item popularity \(s(i)\),
\[
\begin{aligned}
&\gI^-=\{ i : s(i) \leq \rho^- \}, \quad 
\gI^o=\{ i : \rho^- < s(i) < \rho^+ \}\\
&\gI^+=\{ i : s(i) \geq \rho^+ \}
\end{aligned}
\]
where \(\rho^-\) and \(\rho^+\) are popularity thresholds.
We then construct \(N\) sequences \(b_1, \dots, b_N\) by randomly sampling \(T\) items for each sequence from \(\gI^+\) to form \(\gD^+\), and from \(\gI^-\) to form \(\gD^-\).
Thus, \(\gD^+\) captures globally popular content, and \(\gD^-\) represents the long-tail, niche regime. 
For each block \(\ell\) and position \(t\) in the transformer, we record the residual activations \(\vx_{t, \ell}(b)\) for all sequences \(b \in \gD^+ \cup \gD^-\).
We then compute the mean activations of the positive and negative sets:
\[
\vx_{t, \ell}^{+} = \frac{1}{|\gD^+|} \sum_{b \in \gD^+} \vx_{t, \ell}(b),
\qquad
\vx_{t, \ell}^{-} = \frac{1}{|\gD^-|} \sum_{b \in \gD^-} \vx_{t, \ell}(b).
\]
The popularity \emph{steering vector} \(\vv_{t, \ell}\) at block \(\ell\) and position \(t\) is given by the normalized difference in mean activations \citep{arditi2024refusal,hedstrom2025steer},
\begin{equation}
\vv_{t, \ell} = \frac{\tilde{\vv}_{t, \ell}}{\|\tilde{\vv}_{t, \ell}\|} 
\quad \text{with} \quad 
\tilde{\vv}_{t, \ell} = \vx_{t, \ell}^{-} - \vx_{t, \ell}^{+},
\label{eq:steering_vector}
\end{equation}
which identifies the direction in representation space that points away from popular item sequences toward less popular ones.
To reduce popularity at inference time, the steering vector is added to the residual activation of a target sequence,
\begin{equation}
\label{eq:steering}
   \vx_{t, \ell}^{*} = \vx_{t, \ell} + \lambda \, \vv_{t,\ell}, 
\end{equation}
where \(\lambda \in \mathbb{R}\) controls the steering strength.
Not every position in the model encodes popularity equally well. To identify the optimal block \(\ell\) and sequence position \(t\), we train a linear probe \citep{alain2016understanding} to distinguish activations of high-popularity sequences \(\vx_{t,\ell}(b), b \in \gD^+\), from those of low-popularity sequences \(\vx_{t,\ell}(b), b \in \gD^-\). A high linear probe accuracy indicates that popularity differences are well encoded in the activations at block \(\ell\) and position \(t\). In practice, we only steer at the single sequence position  \(t\) and  block \(\ell\) with the highest linear probe accuracy. Typically, this results in selecting the last transformer block at last token position. Please refer to \appref{app:linear_probe} for the linear probe accuracy across blocks and sequence positions. We list implementation details on the sequence generation in \appref{app:generation}.

\subsection{Aligning Popularity via Adaptive Steering Magnitude}
\label{sec:steering_strength}

The steering operation in \Cref{eq:steering} has an important 
limitation: it applies a uniform shift in popularity for all users, 
regardless of whether a user's recommendations are actually 
misaligned with their preferences. This can lead to both 
over-steering and under-steering \citep{hedstrom2025steer}. To address this, we condition the steering correction on each user's estimated popularity bias, allowing both its magnitude and sign to vary across users. Intuitively, if a user’s recommendations are estimated to be more popular than their expressed preference, we steer away from popular items; if they are less popular than preferred, we steer toward more popular items. When no bias is detected, no steering is applied.

Formally, let $e(u)$ denote the popularity bias of user $u$, 
defined as the signed difference between the recommender's and 
the user's empirical popularity quantiles, evaluated at the median 
quantile level $\tau_{\mathrm{med}} = 0.5$:
\begin{equation}
\label{eq:bias}
    e(u) = \hat{\tau}_{\mathrm{med}} - \tau_{\mathrm{med}} \in 
    [-0.5, 0.5],
\end{equation}
where $\tau_{\mathrm{med}} = 0.5$ is fixed and 
$\hat{\tau}_{\mathrm{med}}$ is the empirical fraction of items in 
the user's history with popularity below the recommender's median 
popularity quantile (cf.\ \Cref{eq:tau_hat}). Since 
$\hat{\tau}_{\mathrm{med}} \in [0, 1]$ and $\tau_{\mathrm{med}} = 
0.5$, it follows that $e(u) \in [-0.5, 0.5]$. A negative value 
$e(u) < 0$ indicates that the recommender exhibits a negative bias 
for user $u$ --- the recommended items are less popular than the 
user prefers --- while a positive value $e(u) > 0$ indicates a 
positive bias, meaning recommended items are more popular than 
preferred. Whether $e(u)$ is positive or negative thus determines the sign of the required correction, while its magnitude reflects the severity of misalignment.

To exploit this personalized bias signal, we learn a bias estimator 
$\hat{f} \colon \mathbb{R}^d \to [-0.5, 0.5]$ that predicts $e(u)$ 
from the model's internal activations $\vx_{t,\ell}$. The 
adaptive steering operation then becomes
\begin{equation}
\label{eq:steering_estimator}
    \vx_{t,\ell}^{*} = \vx_{t,\ell}
    + \lambda \, \hat{f}(\vx_{t,\ell}) \, \vv_{t,\ell},
\end{equation}
where $\vv_{t,\ell}$ is the normalized popularity direction 
from \Cref{eq:steering_vector}, and $\lambda$ is again a global
hyperparameter controlling the overall steering strength. 
Note that the sign of $\hat{f}(\vx_{t,\ell})$ determines whether 
the steering vector is added or subtracted, while 
its magnitude scales the correction proportionally to the estimated 
misalignment. We refer to this method as SPREE 
(\textit{Steering PopulaRity toward human prEferEnces}). In practice, we implement $\hat{f}$ as a Lasso 
regression model trained on validation-set activations 
$\vx_{t,\ell}$, using empirically measured biases $e(u)$ as 
targets. Predictive performance is detailed in 
\appref{app:estimator}.

%% file: text/experiments.tex
\section{Experiments}
\label{sec:experiments}

\input{figures/pareto_curves}

We now present the results of our empirical study, in which we evaluate our framework on several benchmark datasets. We compare our proposed method, SPREE, against several baselines, showing that SPREE finds the best trade-off between popularity misalignment (as measured by PCE) and recommendation accuracy (\Cref{sec:exp_pareto}). We also examine calibration curves (\Cref{sec:exp_calibration}), and ablate the impact of SPREE's bias-informed steering (\Cref{sec:exp_ablation}).

\subsection{Experimental Set-up}

\paragraph{Datasets} We evaluate our popularity bias mitigation method on four datasets: Foursquare Tokyo (fs-tky) \citep{yang2014modeling}, 
MovieLens-1M (ml-1m) \citep{movielense}, MovieLens-20M (ml-20m) \citep{movielense}, and RateBeer \citep{mcauley2012learning}. Dataset statistics are summarized in \Cref{tab:dataset_stats}. We employ common data processing practices \citep{he2017neural,tang2018personalized,kang2018self,sun2019bert4rec} and treat each interaction as implicit feedback discarding all items and users that have fewer than 5 interactions. We follow the standard evaluation protocol \citep{he2017neural,he2017translation, tang2018personalized,kang2018self,sun2019bert4rec,rajput2023recommender} and use a leave-one-out evaluation strategy: For all users, the most recent item is used for testing, the second most recent for validation, and the remaining for training. 

\input{tables/datasets_stats}

\paragraph{Base Recommender} 
We use a SASRec \citep{kang2018self} backbone with $L=3$ transformer blocks and a hidden and embedding dimension of $d=64$ across all datasets. The sequence length $T$ is set based on the average user history (see \Cref{tab:dataset_stats}):  
$T=200$ for ml-1m, ml-20m and RateBeer and $T=300$ for Foursquare. All models are trained for 500 epochs using Adam optimizer \citep{kingma2014adam} with a learning rate of 0.001 and batch size of 128. We train three different model seeds per dataset and report averages over seeds. 

\input{figures/calibration}

\paragraph{Baselines} We compare our method to a range of post-hoc approaches designed to mitigate popularity bias. We deliberately focus on inference-time methods, as they operate under the same constraints as SPREE (no training access or model modification), ensuring an apples-to-apples comparison.
The parameter $\alpha \in [0,1]$ controls the adaptation strength: low values keep the predictions close to the base model, while higher values apply stronger mitigation.
\begin{itemize}[leftmargin=*]
    \item \textbf{Base}: the unadapted, base sequential recommender (SASRec).

    \item \textbf{Inverse Popularity Ranking (IPR)} \cite{zhang2010niche}: 
    This baseline downgrades the logits of the base recommender according to the item's popularity, 
    $\logits^{\text{IPR}}_{i} = \frac{\logits^{\text{Base}}_{i}}{1 + \alpha \times s^{\text{IPR}}(i)}$ 
    with $s^{\text{IPR}}(i) = s(i)/\max_{i \in \mathcal{I}} s(i)$ denoting normalized raw popularity. 
    As a consequence, popular items become less likely to be recommended.

    \item \textbf{Personalized Popularity (PP)} \citep{abbattista2024enhancing}: 
    This method adapts logits at inference time by interpolating the logits of the base recommender 
    $\logits^{\text{Base}}_{i}$ 
    with a user-specific popularity score $s^{\text{PP}}(i,u)$ that ranks item $i$ based on the number of past interactions of user $u$, 
    $\logits^{\text{PP}}_{i} 
        = \alpha\, s^{\text{PP}}(i) 
        + (1 - \alpha)\, \logits^{\text{Base}}_{i}$.
    The method boosts items the user has previously interacted with.

    \item \textbf{Random Neighbors} \citep{ahmadov2025opening}: This baseline randomly samples $K$ items from the neighborhood of the user embedding $\vh$. The neighborhood is defined as the $M = K(1+\alpha)$ closest items to $\vh$. This method evaluates whether a better popularity--accuracy trade-off exists in the vicinity of the base recommender's top predictions.
    
    \item \textbf{PopSteer} \citep{ahmadov2025opening}: A steering-based method like ours, PopSteer uses sparse auto-encoders (SAE) \citep{cunningham2023sparse} to identify and switch off neurons in the user embedding that encode popularity. Please refer to \appref{app:popsteer} for implementation details. 
\end{itemize}

\subsection{Alignment-Performance Trade-off}
\label{sec:exp_pareto}
In our first experiment, we evaluate the effectiveness of SPREE across datasets. Specifically, we study how well SPREE reduces misalignment between the recommender’s and users’ popularity distributions, measured by PCE@100, while preserving predictive performance, measured by NDCG@100. We vary the mitigation intensity of SPREE via the global steering strength $\lambda \in \{0, 1, 2, 4, 8, 16, 32\}$. For PopSteer, we use steering strengths $\lambda \in \{0, 0.1, 0.2, \dots, 0.9, 1\}$; and for IPR, random neighbors and PP we vary $\alpha \in \{0, 0.1, \dots, 0.9, 1\}$.

\Cref{fig:pareto_curves} shows Pareto curves of PCE@100 versus NDCG@100 across all datasets. We observe that only PP (\redline) and SPREE (\greenline) consistently reduce PCE@100 relative to the SASRec baseline. This is intuitive, as both methods align recommendations to the user's history. However, PP leads to a steady degradation in NDCG@100 on all datasets except for fs-tky. This is because promoting items a user has already interacted with (as PP does) helps on highly repetitive sequences such as the foursquare check-in data but fails in recommending novel items. In contrast, SPREE largely preserves predictive performance on fs-tky and ml-1m and even yields a slight improvement on RateBeer. Methods that globally reduce popularity for all users (IPR and PopSteer) result in moderate NDCG@100 decrease but fail to improve user alignment, as indicated by increased PCE@100.

\subsection{Calibration Assessment}
\label{sec:exp_calibration}
Next, we more closely assess the alignment of the recommender’s popularity distribution with the users’ historical popularity preferences. We are particularly interested in three aspects. (i) The direction of misalignment: Are recommended items systematically too popular (positive bias) or too niche (negative bias)? (ii) The region of misalignment in the user's popularity distribution: Does misalignment concern the lower tail, body, or upper tail of the recommended items? Lastly, we are interested in (iii) the effect of mitigation methods on the recommender's popularity distribution. 
We use the calibration curve (\Cref{sec:diagnostic}) to examine (i) - (iii). Unlike PCE, the calibration curve allows to infer the direction of the bias (D1 ordinality) for each region in the recommender's popularity distribution (i.e for each quantile $\tau \in \{0, 0.1, \dots 1\}$). 
We run each method with the highest mitigation strength ($\alpha=1$ for IPR, random neighbors and PP; $\lambda = 1$ for PopSteer and $\lambda=32$ for SPREE) and compute the average calibration curve across all users.

\Cref{fig:calib} depicts calibration curves. Regarding (i), we note that the base recommender is well calibrated on fs-tky, but exhibits a predominantly negative bias  for ml-1m, ml-20m and RateBeer (the black line (\blackline) is below the perfectly calibrated oracle (\graydashedline)). For example, at the recommender’s median popularity quantile $\tau =0.5$, the corresponding user popularity quantile is lower ($\hat{\tau} \approx 0.45$). This indicates that items recommended as moderately popular by the model fall among the less popular items of the users’ historical interaction distributions, suggesting the average user prefers more popular items. 
Regarding (ii) location of misalignment, the negative bias is most pronounced in the body and upper tail of the recommended items, while the lower tail (\(\tau < 0.4\)) exhibits little to slightly positive bias, indicating that already niche recommendations could be even less popular. This pattern suggests that the recommender’s popularity distribution is overly concentrated and lacks sufficient spread.
Turning to (iii) mitigation effects, we observe that methods that uniformly reduce popularity, i.e.  IPR (\cyanline), random neighbors (\blueline) and popsteer (\purpleline) tend to amplify the negative bias, leading to worse user calibration. In contrast, the PP baseline (\redline) can oversample users’ most popular items, resulting in a consistent positive bias across all quantiles on ml-1m and ml-20m. Our method (\greenline) instead shifts the calibration curve closer to the perfectly calibrated oracle, improving alignment without overcorrecting.

\subsection{Ablation of Bias-conditioned Steering Strength}
\label{sec:exp_ablation}

Next, we assess the impact of the bias-informed steering strength (\Cref{sec:steering_strength}) on alignment effectiveness. SPREE employs a bias estimator to predict both the severity and direction of popularity bias on a per-user basis and scales the steering magnitude accordingly. As an ablation, we consider a \emph{vanilla} variant that applies the same steering strength to all users, globally steering recommendations away from popular items (\Cref{eq:steering}). Intuitively, such uniform steering is expected to reduce overall popularity as defined in systematized concept (S1) (\Cref{sec:existing_metrics}) but may not necessarily improve alignment with individual user preferences as defined in (S3).
To evaluate this, we compare SPREE with the bias estimator (\Cref{eq:steering_estimator}) to SPREE without the estimator (\textsc{SPREE w/o}~$\hat{f}$; \Cref{eq:steering}). The steering direction $\vv_{t,\ell}$, as well as the intervention block $\ell$ and position $t$, are identical across both variants. We use PCE@100 to measure user-centric popularity bias (S3) and ALRP@100 \citep{lichtenberglarge} to quantify the absolute popularity level of the recommender system (S1).

\Cref{tab:ablation} reports PCE@100 and ALRP@100 for the largest steering strength $\lambda \in \{1, 2, 4, 8, 16, 32\}$ that results in at most a $10\%$ reduction in NDCG@100. The bias-informed variant consistently reduces PCE@100 across all datasets, indicating improved user-level alignment. In contrast, the vanilla variant improves alignment only on ml-20m, while substantially worsening alignment on other datasets—by more than $100\%$ on fs-tky and ml-1m. Examining ALRP@100, we observe that the vanilla variant indeed reduces global popularity across datasets. By contrast, SPREE maintains a roughly constant popularity level, suggesting that it redistributes popularity across users rather than shifting it globally—effectively resulting in a zero-sum adjustment.

\input{tables/ablation}

%% file: figures/pareto_curves.tex
\begin{figure*}[h!]
    \centering
    \scalebox{0.9}{
        \includegraphics[width=\linewidth]{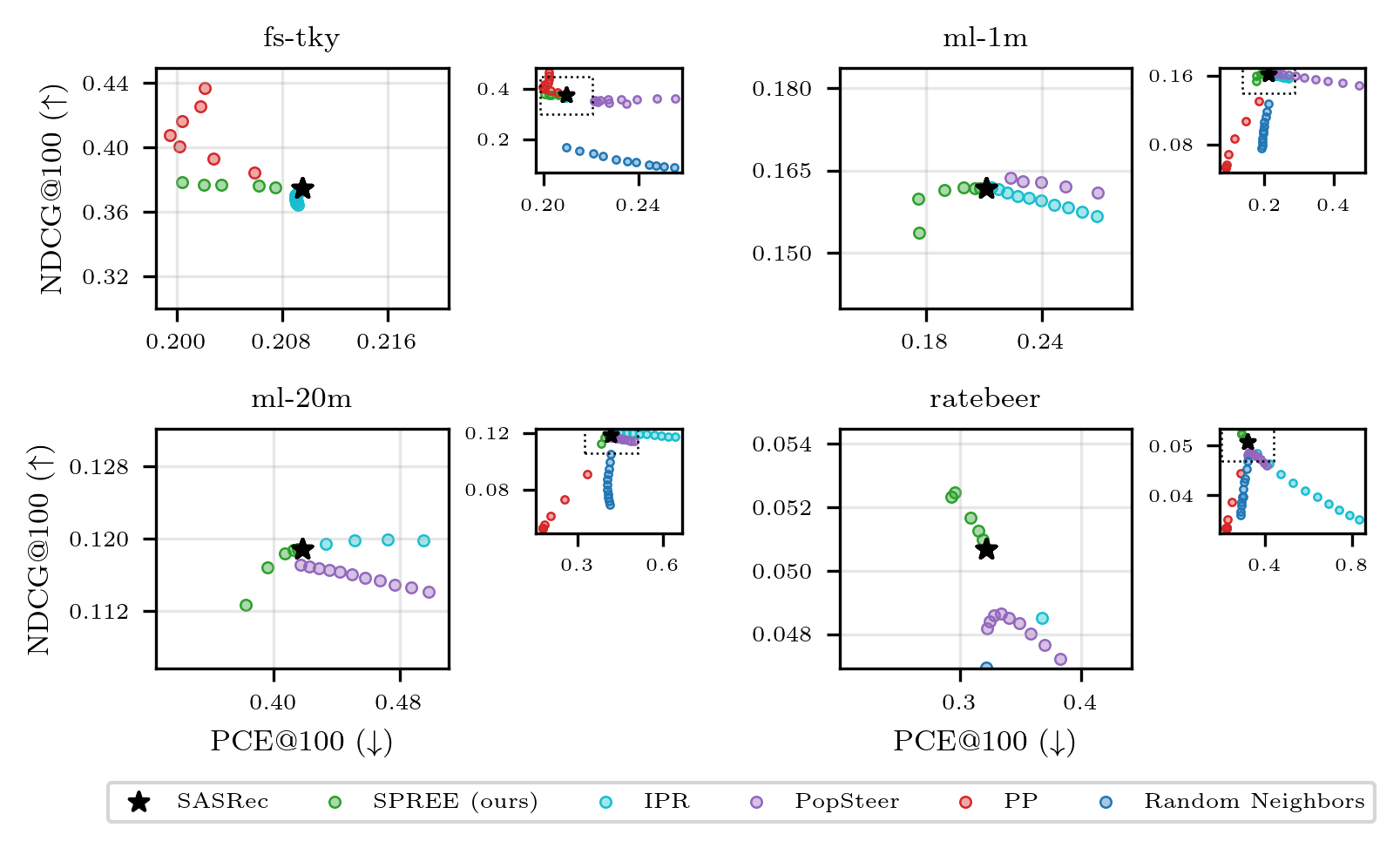}
    }
    \caption{Trade-off between performance (NDCG) and popularity alignment (PCE) across different datasets and mitigation baselines. Points represent different mitigation strengths. The plots show the vicinity of the base model (larger plots on the left) and the global trends (smaller plots on the right).}
    \label{fig:pareto_curves}
\end{figure*}

%% file: tables/datasets_stats.tex
\begin{table}[h!]
\centering
\caption{Dataset statistics}
\label{tab:dataset_stats}
\scalebox{0.85}{
\begin{tabular}{lrrrrr}
\toprule
\multirow{2}{*}{\textbf{Dataset}} & \multirow{2}{*}{\textbf{\#users}} & \multirow{2}{*}{\textbf{\#items}} & \multirow{2}{*}{\textbf{\#interact.}} & \textbf{ interact.} & \textbf{interact.} \\
 & & & & \textbf{per user} & \textbf{per item} \\
\midrule
Foursquare      &  2,293 &  15,177 &  494,807 & 216 & 33  \\
MovieLens-1m     & 6,040   & 3,416   &  999,611 & 166 & 292 \\
MovieLens-20m     & 138,493   & 18,345  & 19,984,024 & 144 & 1089 \\
RateBeer & 12,221 & 53,674 & 2,631,157 & 215 & 49 \\
\bottomrule
\end{tabular}
}
\end{table}

%% file: figures/calibration.tex
\begin{figure*}
    \centering
    \includegraphics[width=\linewidth]{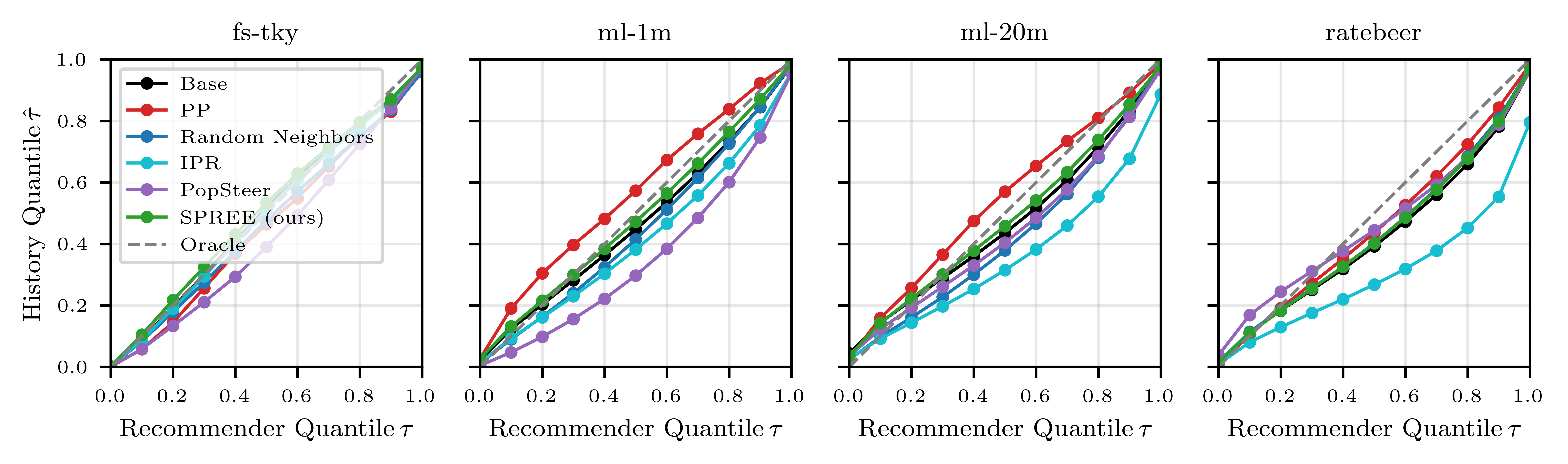}
    \caption{Average calibration curves across users. Except for fs-tky, where the base model is already well calibrated, the recommender tends to favor less popular items (black curve below the dashed gray identity line). SPREE (green) mitigates this bias by bringing the curve closer to the oracle identity. In contrast, baselines that reduce global popularity (PopSteer and IPR) increase this miscalibration, while Personalized Popularity (PP) may overcorrect by recommending only the most popular items in the user’s history.}
    \label{fig:calib}
\end{figure*}

%% file: tables/ablation.tex
\begin{table}[t!]

\centering
\caption{Ablation of the steering strength based on each user’s estimated bias (\Cref{eq:steering_estimator}): SPREE without an estimator (\Cref{eq:steering}) reduces global popularity (S1) but may increase popularity alignment (S3). Using a bias estimator improves alignment while keeping global popularity roughly constant.}

\label{tab:ablation}
\scalebox{0.75}{
\begin{tabular}{lc|cc||c|cc}
\toprule
Dataset
& \multicolumn{3}{c}{PCE@100 $(\downarrow)$} 
& \multicolumn{3}{c}{ALRP@100} \\
\cmidrule(lr){2-4} \cmidrule(lr){5-7}
& Base  & SPREE  & SPREE w/o $\hat{f}$ 
& Base  & SPREE  & SPREE w/o $\hat{f}$ \\
\midrule
fs-tky 
& 0.210 
& 0.200 \textcolor{TabBlue}{(-4\%)} 
& 0.509 \textcolor{TabOrange}{(+143\%)} 
& 4.663 
& 4.713 \textcolor{TabOrange}{(+1\%)} 
& 3.636 \textcolor{TabBlue}{(-22\%)} \\

ml-1m 
& 0.211 
& 0.176 \textcolor{TabBlue}{(-16\%)} 
& 0.434 \textcolor{TabOrange}{(+106\%)} 
& 6.339 
& 6.337 \textcolor{TabBlue}{(-0\%)} 
& 5.883 \textcolor{TabBlue}{(-7\%)} \\

ml-20m 
& 0.418 
& 0.382 \textcolor{TabBlue}{(-9\%)} 
& 0.378 \textcolor{TabBlue}{(-10\%)} 
& 9.012 
& 9.017 \textcolor{TabOrange}{(+0\%)} 
& 8.912 \textcolor{TabBlue}{(-1\%)} \\

ratebeer 
& 0.321 
& 0.293 \textcolor{TabBlue}{(-9\%)} 
& 0.335 \textcolor{TabOrange}{(+4\%)} 
& 5.652 
& 5.860 \textcolor{TabOrange}{(+4\%)} 
& 5.577 \textcolor{TabBlue}{(-1\%)} \\

\bottomrule
\end{tabular}
}

\end{table}

%% file: text/conclusion.tex
\section{Conclusion}
In this work, we focus on popularity bias as a user–recommender alignment problem. By assessing existing metrics, we derive desiderata for measuring such misalignment validly. We introduce a calibration perspective on user-centric popularity bias and propose a new metric, PCE, to assess it. To mitigate misalignment, we propose SPREE, an inference-time activation steering method that aligns the model’s internal representations toward estimated human preferences. We show that SPREE consistently improves user alignment while largely preserving predictive performance.

\paragraph{Limitations and Future Work.}
While we have shown that SPREE can improve popularity alignment between recommenders and users, the degree of improvement is limited by the structure of the item embedding space: steering is only effective if locally popularity-calibrated regions exist since retrieval is based on nearest neighbors. Moreover, our method targets bias around the median popularity, which restricts control to shifting the central tendency of the popularity distribution. Future work could target specific parts of the distribution (e.g., the lower or upper tails). Finally, extending this framework beyond popularity to other alignment objectives, such as genre diversity \citep{zhao2025fairness} or exposure to minority content \citep{khan2023effects}, is an interesting avenue for future research.

%% file: text/appendix/appendix.tex
\section{Experimental Details}
\subsection{Generation of Contrastive Datasets}
\label{app:generation}

To elicit the steering vector and identify the optimal steering position, we sample $|\gD^+|=5000$ artificial mainstream users and $|\gD^-|=5000$ artificial tail users. Sequence lengths are matched to each dataset’s context window, reserving the first 100 positions for padding: we sample 100 items for ml-1m, ml-20m, and RateBeer, and 200 items for fs-tky. Popularity thresholds $\rho^+$ and $\rho^-$ are defined as the popularity score $s$ corresponding to the top 10\% and bottom 10\% most popular items, respectively.

\subsection{Implementation Details for PopSteer}
\label{app:popsteer}
Following \cite{ahmadov2025opening}, we train a sparse autoencoder (SAE) on the final-position user representations extracted from the frozen, pre-trained SASRec models. Further following \cite{ahmadov2025opening}, we inspected SAE dimensions of [512, 1024, 2048, 4096] and found 512 to be best performing for all datasets. The sparsity parameter is 32, with a learning 
rate of 1e-04. We set the maximum epochs to 500 with an early stopping patience of 10 and a 90/10 train-validation split for all datasets. All models stopped training prior to the maximum set epochs. SAEs are trained for each seed of each trained SASRec model independently.

\section{Results on Steering Efficacy}
\subsection{Popularity Encoded in the Model: Linear Probe Accuracy}
\label{app:linear_probe}

To verify that popularity is encoded in the model’s internal representations, we train a linear probe to distinguish between activations of sequences in $\gD^+$ (popular items) and $\gD^-$ (tail items). We zero-pad the first 100 sequence positions of the model to isolate the effect of subsequent tokens.

\Cref{fig:linear_probe} reports the probe accuracy on a held-out set of artificial users. Immediately after the padded region, the accuracy rises to nearly 100\%, indicating that popularity is linearly separable in activation space. Moreover, later blocks (i.e. block 3) consistently achieve higher probe accuracy than earlier blocks (i.e. block 1), suggesting that popularity information becomes more explicitly represented in deeper layers. For the large ml-20m dataset, all blocks converge to almost perfect separability, whereas for the smaller fs-tky dataset, the differences between blocks are more pronounced.

\input{figures/linear_probe_accuracy}

\subsection{Bias Encoded in the Model: Bias Estimator Performance}
\label{app:estimator}

To verify that the bias, rather than popularity itself, is encoded in the model’s representations, we evaluate the estimator’s ability to predict user-level popularity misalignment from internal activations. As described in \Cref{sec:steering_strength}, SPREE employs a bias estimator that predicts the difference in quantiles at the popularity score corresponding to the median of the recommendations. We split the validation set into disjoint training and test subsets, train a Lasso regression on the former, and evaluate it on the held-out set. \Cref{tab:estimator_performance} reports performance in terms of $R^2$ and test MSE.

The results mirror the pattern observed for the linear probe accuracy in \Cref{app:linear_probe}. Performance is strongest on the large ml-20m dataset, where the estimator reaches high $R^2$ and low MSE, indicating that user-level popularity bias is reliably encoded in the activations. In contrast, predictability is weakest on the smallest, fs-tky, suggesting that bias signals are harder to recover from the activations.

\input{tables/estimator_error}

\newpage

%% file: figures/linear_probe_accuracy.tex
\begin{figure}[h!]
    \centering
    \includegraphics[width=1\linewidth]{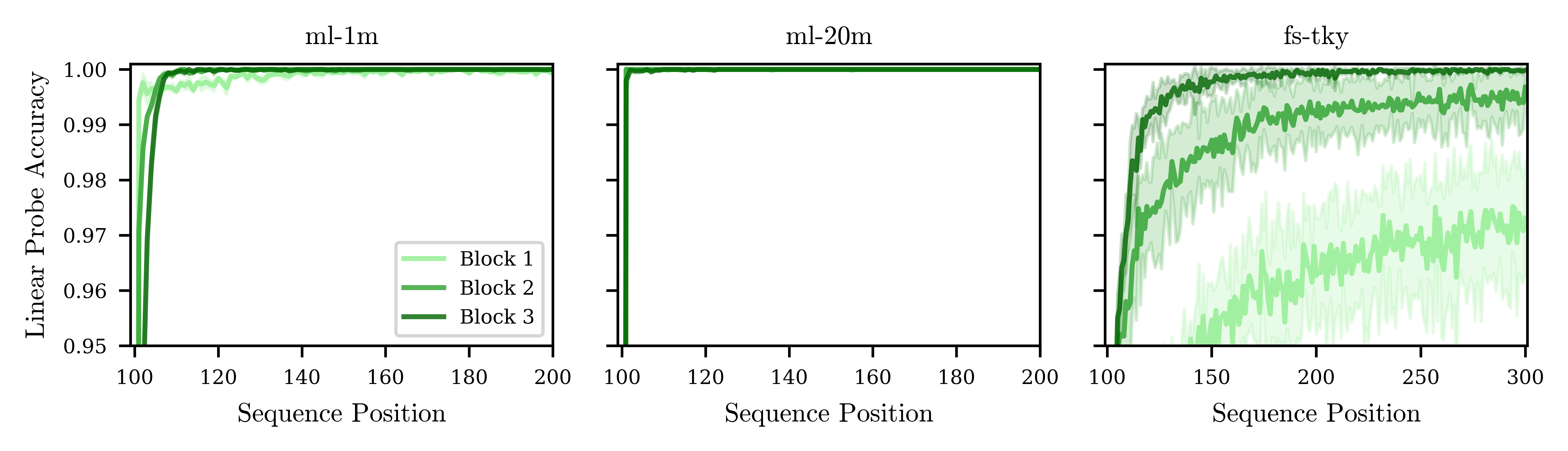}
    \caption{Linear probe accuracy across sequence positions $t$ and transformer blocks $l$ for ml-1m, ml-20m, and fs-tky. After the padded region, accuracy rapidly converges toward 100\%, indicating that popularity is linearly separable in activation space. Popularity is most strongly encoded at the final sequence positions in the last block.}

    \label{fig:linear_probe}
\end{figure}

%% file: tables/estimator_error.tex
\begin{table}[h!]
\centering
\caption{Predictive performance of the bias estimator $\hat{f}$ on a hold-out set of user activations.}
\label{tab:estimator_performance}
\begin{tabular}{lcc}
\toprule
Dataset   & MSE & $R^2$ \\
\midrule
fs-tky    & $0.027 \pm 0.002$ & $0.182 \pm 0.042$ \\
ml-1m     & $0.015 \pm 0.003$ & $0.425 \pm 0.100$ \\
ml-20m    & $0.019 \pm 0.002$ & $0.638 \pm 0.029$ \\
ratebeer  & $0.026 \pm 0.000$ & $0.223 \pm 0.031$ \\
\bottomrule
\end{tabular}
\end{table}